\documentstyle[11pt,newpasp,twoside]{article}
\def\edcomment#1{\iffalse\marginpar{\raggedright\sl#1\/}\else\relax\fi}
\reversemarginpar

\begin{document}
\title{Light Curve Analysis of Be Star Candidates in the Small
Magellanic Cloud}
 \author{R.E. Mennickent, G. Pietrzynski, W. Gieren}
\affil{Departamento de F\'{\i}sica, Universidad de Concepci\'on,
Casilla 160-C, Concepci\'on, Chile.}
%\author{Ima Co-Author}
%\affil{The Name of My Institution, The Full Address of My Institution}

\begin{abstract}
The recent discovery,  based on a search of public OGLE 
data, of Be star candidates in the Small Magellanic Cloud
showing spectacular photometric variations (Mennickent et al. 2002),
has motivated further analysis  of their light curves. Here we
report the results of a statistical study of the light curves of
Type-1, Type-2 and Type-3 stars. Type-3 stars show a
bimodal period distribution, with a main, broad peak, between 20
and 130 days, and a narrower, secondary peak, between 140 and 210 days.
We also find that, among Type-3 stars, the maximum amplitude of
oscillation correlates with the
system luminosity, in the sense that only low luminosity
systems show large amplitude oscillations.
In general, the amplitude of these oscillations tends to increase
with the photometric period.  A parametric study shows no
correlation between the  amplitude, duration and asymmetry
of Type-1 outbursts, and amplitude of Type-2 jumps, with the
available stellar photometric parameters.

\end{abstract}

\section{Introduction}

Detailed  studies of Be stars in the Small Magellanic Cloud (SMC) have been
performed only in recent years, being especially
confined to open clusters like
NGC 330 (e.g. Keller, Wood and Bessell 1999).
These studies show the importance of studying Be stars in
the low metallicity environment of the SMC
since they serve as probes to test for the mechanisms of
disk formation and of global disk oscillations
(Baade et al.\ 2001, Hummel et al.\ 1999).

Recently, Mennickent et al. (2002, hereafter M02), based on a search of public OGLE 
data, reported the discovery of
about 1000 Be star candidates in the SMC showing spectacular light curve variations.
M02 classified their sample in Type-1 stars (showing outbursts), Type-2 stars
(showing sudden luminosity jumps), Type-3 stars (showing periodic or near-periodic
variations) and Type-4 stars (showing similar light curves that Galactic
Be stars). M02 present examples of light curves for each one of these groups,
giving some basic statistical information about colours and periodicities.
Here we pursue this analysis, with a more detailed
statistical study of the photometric properties of each group.

\section{Results}

\subsection{Type-1 Stars}

We have investigated the (I-band) outburst properties of a sample
of these stars (the first 40 stars in the $RA$ ordered list,
corresponding to 30\% of the total), measuring  the
amplitude $A$, duration $\Delta T$ and asymmetry $\delta$, of all detected
outbursts, amounting to 105, according to the definitions shown in Fig.\,1.
The $A$ distribution reveals
a strong, broad peak with a maximum around 0.1 mag (Fig.\,2).
The $\Delta T$ distribution shows a maximum around 25 d, with a
steep decrease toward shorter periods and a much flatter decrease
towards longer periods. The $\delta$ distribution reveals that
about 30\% of the outbursts have $\delta$ $<$ 1, the rest shows
rapid rises and slower declines.

\subsection{Type-2 Stars}

We measured the amplitude of the photometric jumps in the I-band
for these stars.
Few objects showed two or even three (generally low amplitude)
jumps during the observing period, the more prominent jump was used for
the statistics in these cases. 
The histogram of amplitudes
in Fig.\,3 shows a prominent concentration around 0.15$-$0.25 mag.
There is no correlation between the jump amplitude and the
stellar magnitude, $B-V$ or $V-I$ color.

\subsection{Type-3 Stars}

Some information about the variability of these stars is already
given in M02. Here we focus on the
amplitudes of these variables, the morphology
of their I-band light curves and the overall
period distribution. Accordingly to Table-5 of M02,
Type-3 stars could be classified as follows: those showing
a possible period change (2 cases),
those showing eclipses (5 cases), those whose periodicity appears
after removing a long-term tendency (17 cases),
and those showing rather clear oscillations, although usually
of variable amplitude (54 cases). We fit the phase curves
of the two later groups (obtained using residuals when neccessary)
with a simple sine function, obtaining the amplitudes for the
oscillations. The period distribution for Type-3 stars shown in
Fig.\,4. suggests the existence of a bimodal distribution. We
observe a broad concentration of stars with periods
between 20 and 130 days and a narrower concentration with periods
between 140 and 210 days. There is no priviliged location for the
different Type-3 classes in the above diagram (neither in the
diagrams shown later in this section).  Both bright and faint stars
are found inside the two peaks of the period distribution. Surprisingly,
we find that
all stars in the secondary peak show rather ``clean" light curves; 
for these stars it was not neccessary to remove a 
a long-term trend in order to find the period, as occurred for 
17 Type-3
stars (M02). Some phase curves
of Type-3 stars with periods in the range of 140 to 210 days are shown in Fig.\,5. 
Another interesting result is
that large amplitude oscillations are
only observed in low luminosity systems (Fig.\,6).
On the other hand, the amplitude$-$period
diagram (Fig.\,6) shows that, in general,
the oscillation amplitude tends to increase with the period. It is worthy
to mention that inclination effects are not included in Fig.\,7, and they could
be responsible for the large scatter shown by the data.

\section{Discussion}

The first impression, after the visual inspection of Type-1 light curves,
was that the outbursts were divided in hump-like and sharp outbursts. However,
a clear separation between these groups is not confirmed by the distributions of the
parameters $A$, $\Delta T$ and $\delta$, which instead suggest smooth transitions between
one outburst type and the other. The above results do not help to discriminate between the
possible hypotheses for these outbursts rised by M02, viz.\, thermal instabilites in
circumstellar discs or accretion by an unseen white dwarf.

The discovery of a bimodal period
distribution in Type-3 stars is surprising
and deserves further study. At present,
we do not have an explanation for this finding, but we
expect to make a deeper analysis of these stars incorporating spectroscopic
data at the end of this year.

The fact that {\it only} Type-3 systems with faint absolute magnitude show large amplitude
oscillations
could be the clue to understand the underlying cause of this phenomenon.
M02 suggested that Type-3 variability could be due to some kind of oscillation
in a circumstellar envelope. More massive envelopes could trigger larger amplitude
oscillations.
Perhaps low luminosity stars can host massive envelopes, whereas the higher
radiation pressure of higher luminosity stars conspires against the formation and maintenance
of massive envelopes. This could explain the absence of large amplitude variations in high luminosity
systems.

\acknowledgments

We thank Ligia Barros and Paulina Assmann for
their help during the process of light curve
analysis. This research was supported by Grant Fondecyt
1000324.

\section{References}

Baade, D., Rivinius, T., {\v S}tefl, S., \& Kaufer, A., 2002, \aap, 383,
L31.\\
Hummel, W., G\"{a}ssler, W.,
Muschielok, B.,  Schink, H.,
 Nicklas, H., et al.\ 2001, \aap, 371, 932.\\
Keller, S.~C., Wood, P.~R., \& Bessell, M.~S., 1999, \aaps, 134, 489  \\
Mennickent, R.E., Pietrzy\'nski, G., Gieren, W., \& Szewczyk, O., 2002, \aap, in press.\\

{\bf Figure Captions}

Fig.1: Definition of amplitude (A), duration ($\Delta T$) and
asymmetry ($\delta$)
for Type-1 star outbursts.

Fig.2: Distributions for the parameters A, $\Delta T$ and $\delta$.

Fig.3: The distribution of amplitudes for Type-2 star jumps.

Fig.4: The period histogram for Type-3 stars. Note the bimodal distribution.

Fig.5: Examples of phase curves for Type-3 stars with periods between 140 and 210 days.

Fig.6: The amplitude$-$magnitude diagram for Type-3 stars. Error bars are plotted when available.

Fig.7: The amplitude$-$period diagram for Type-3 stars.

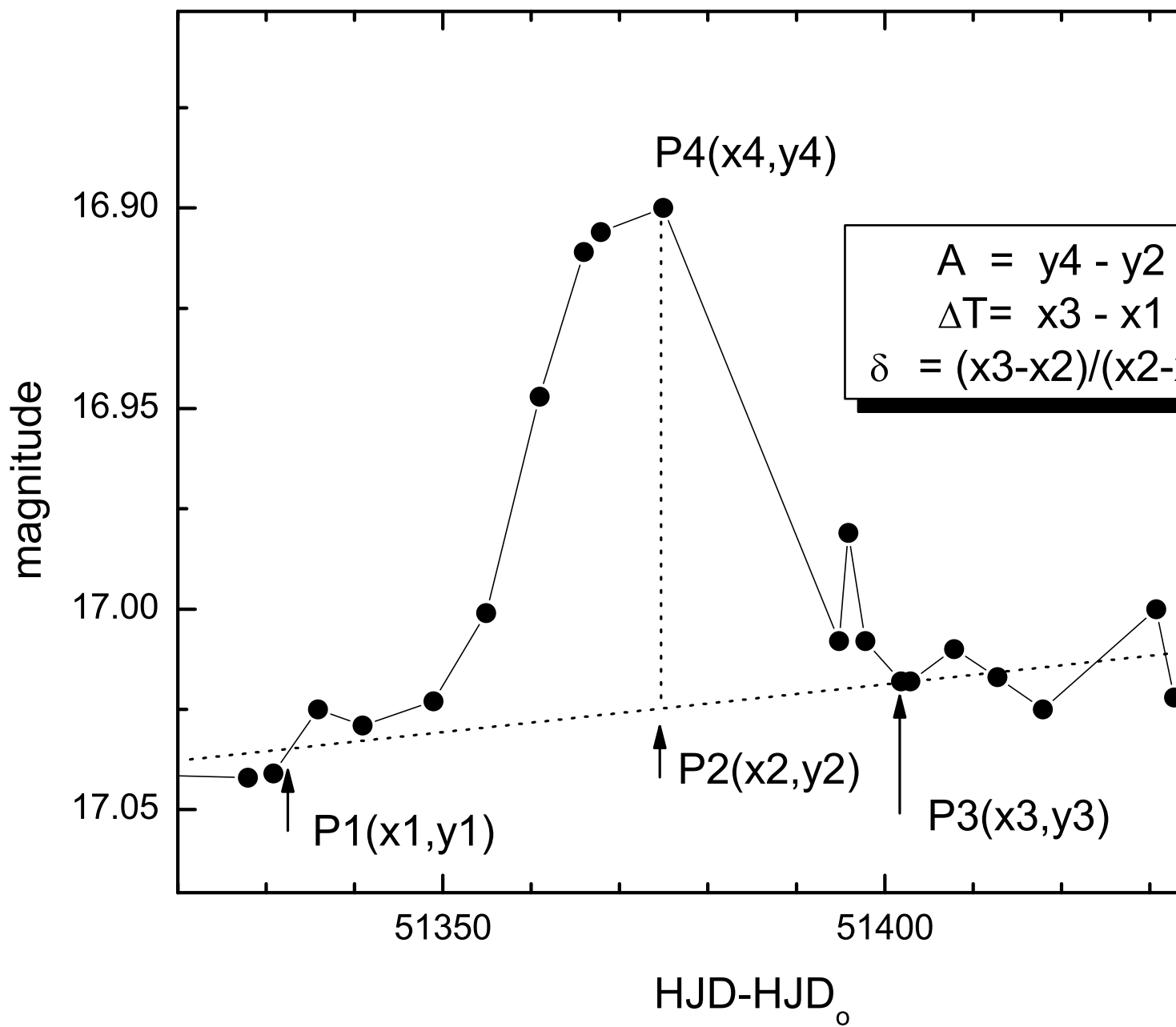
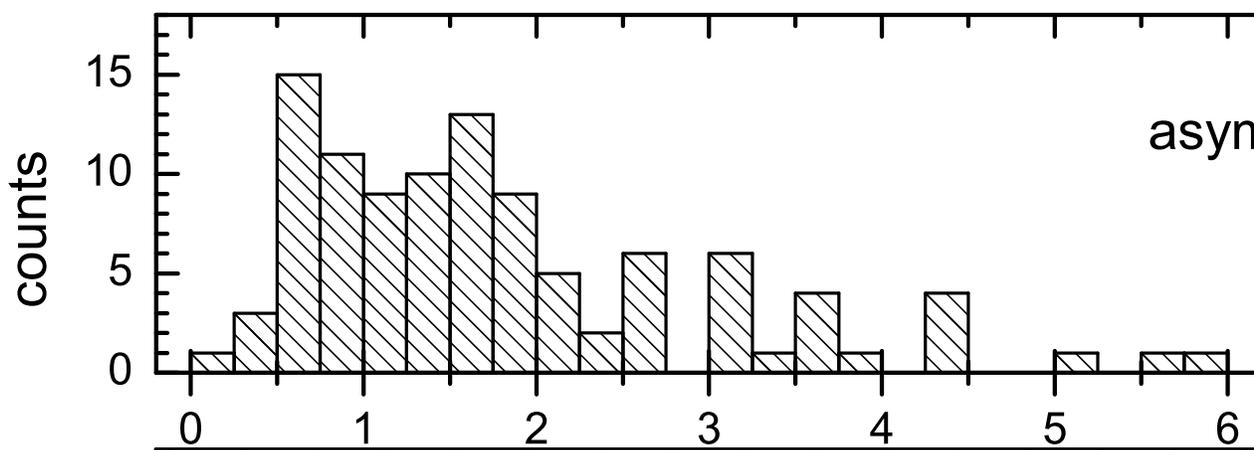
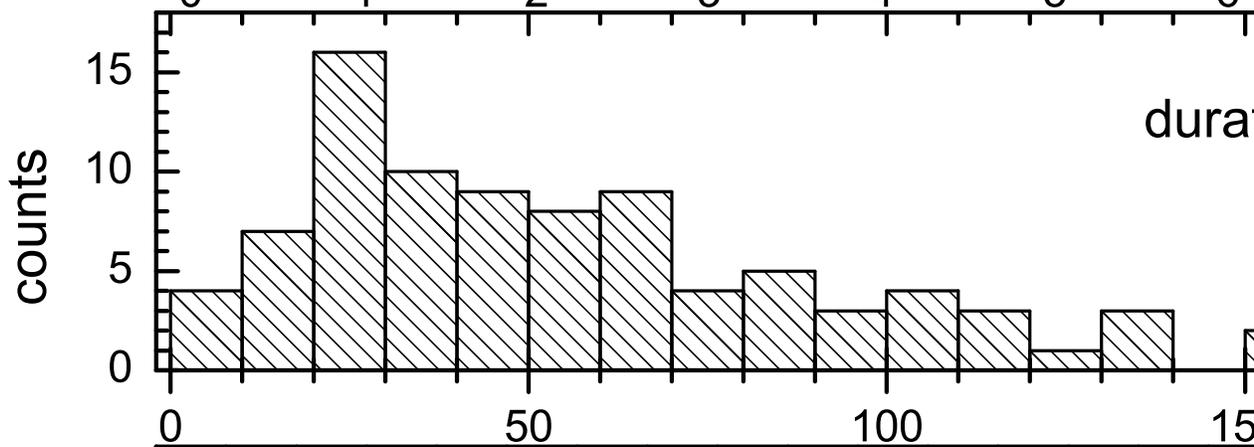
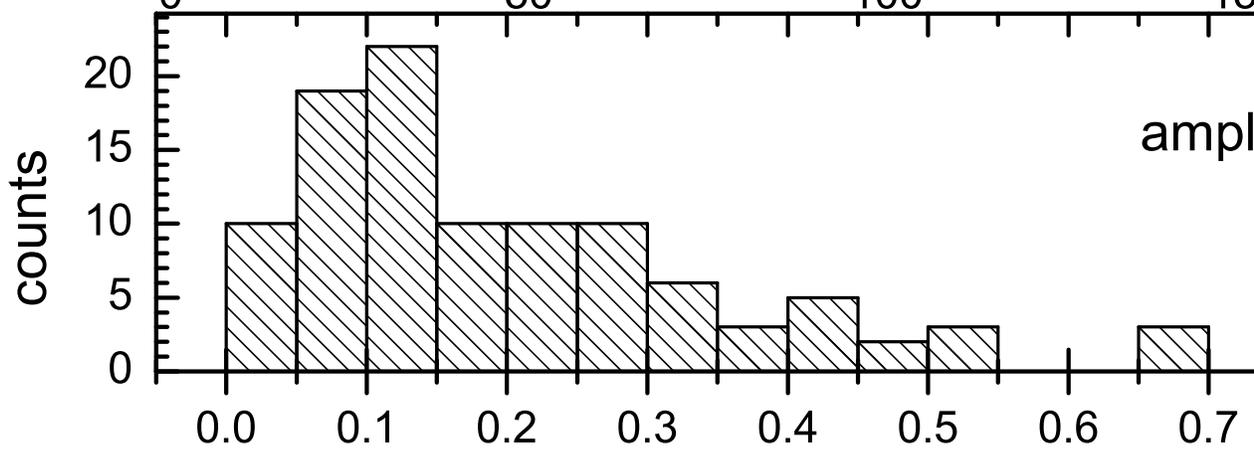
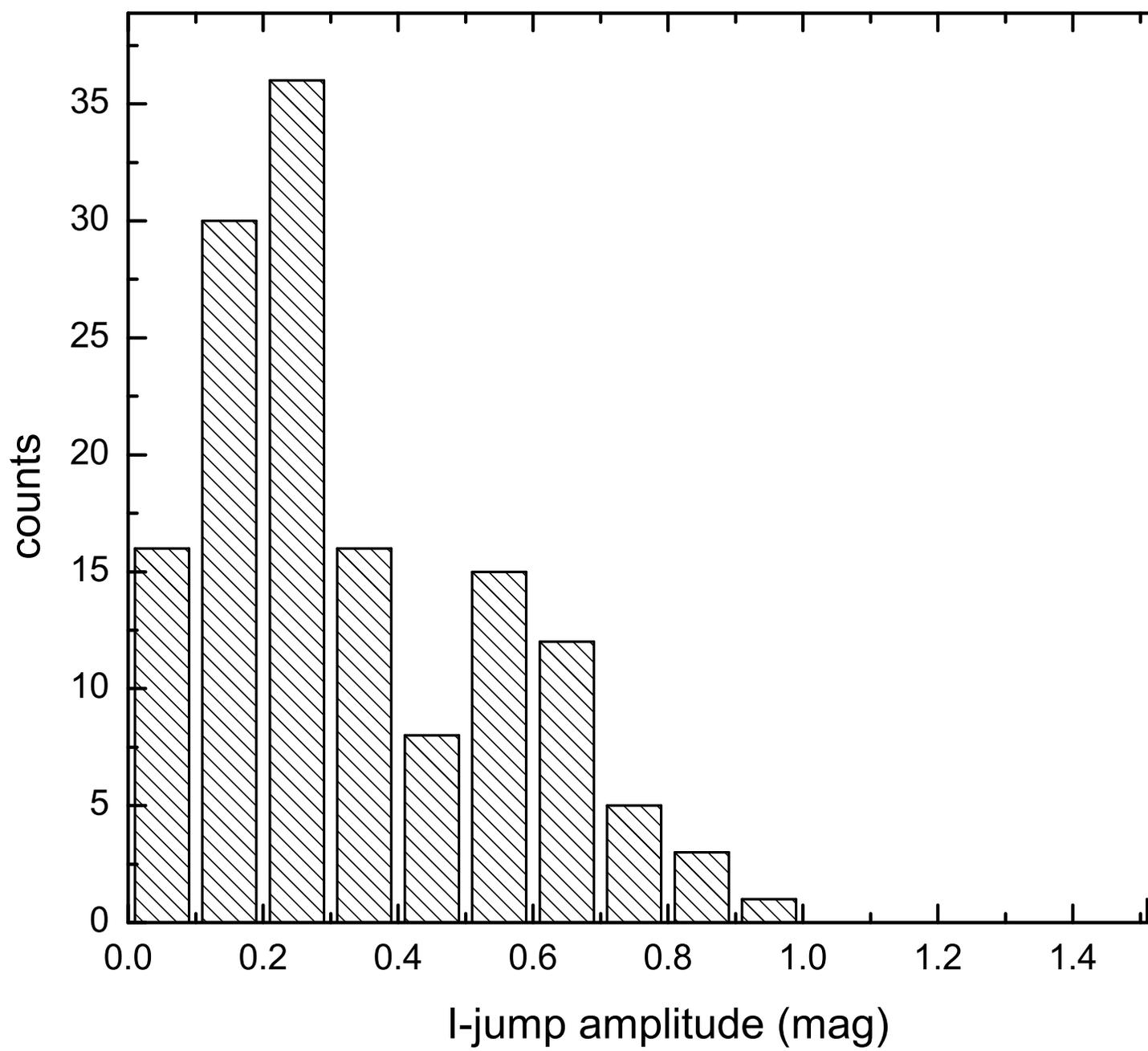
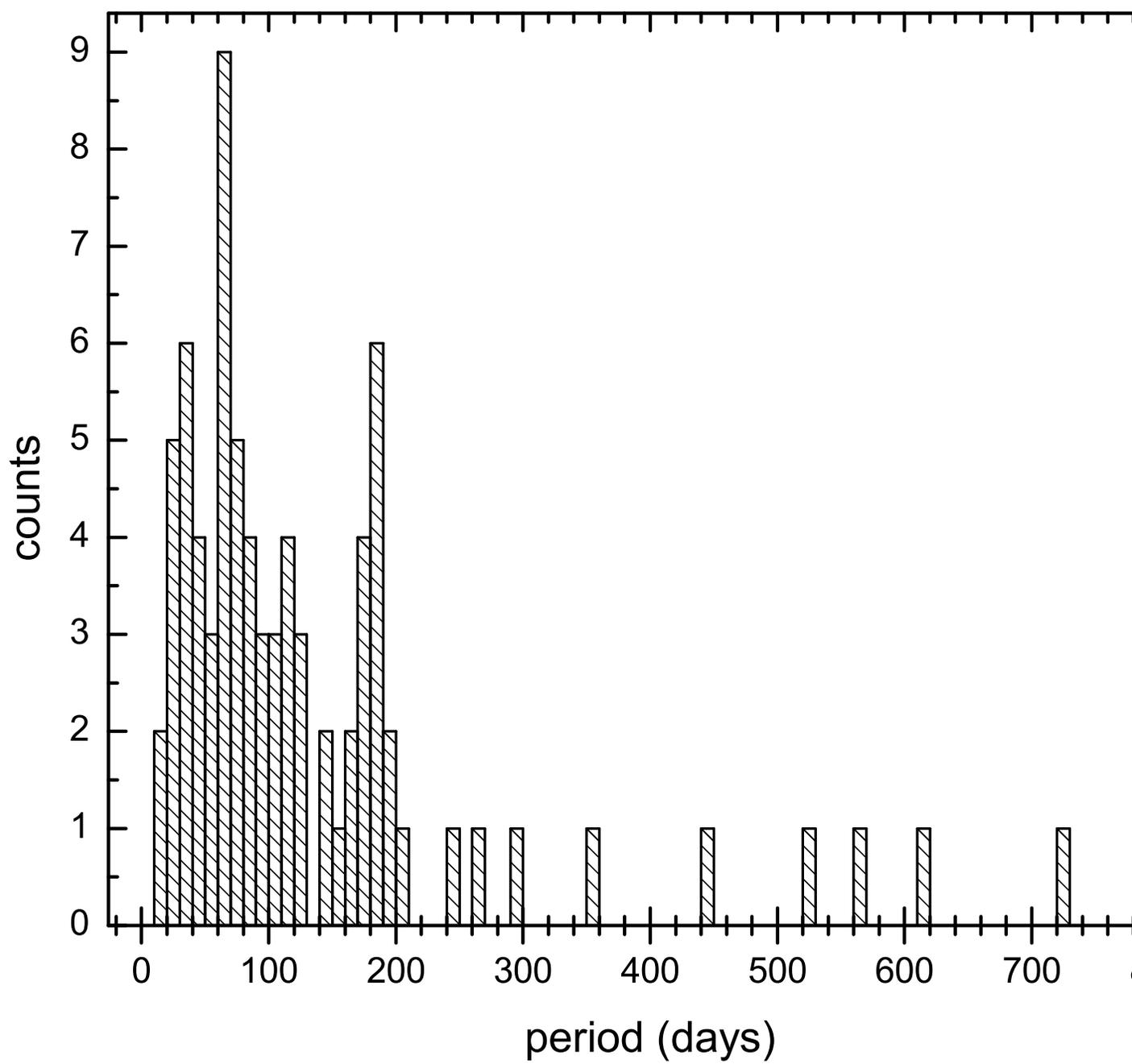
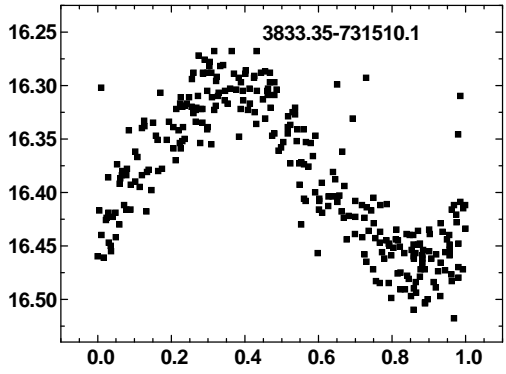
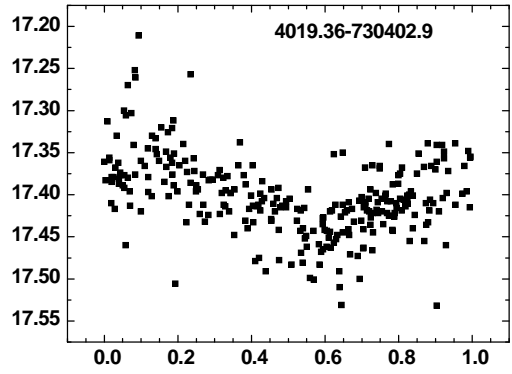
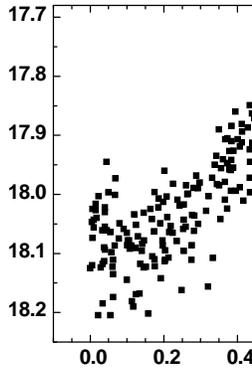
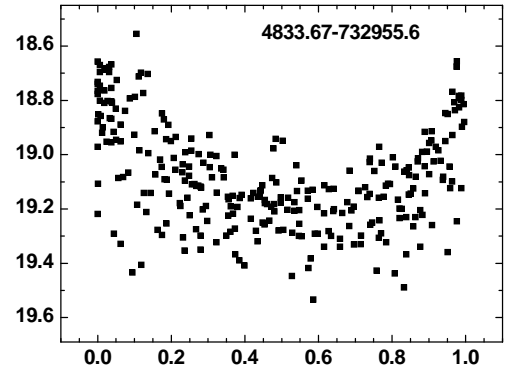
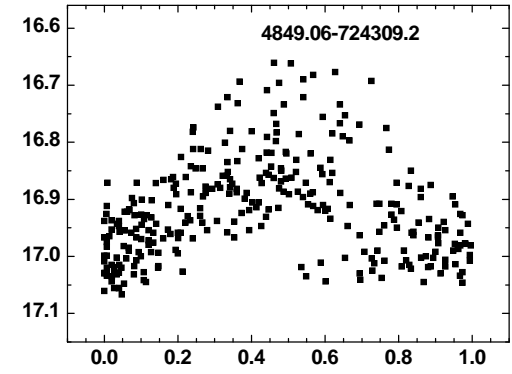
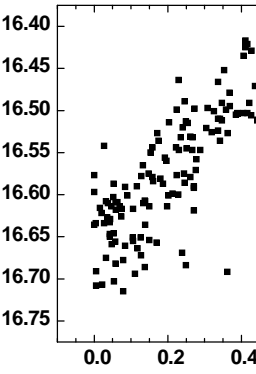
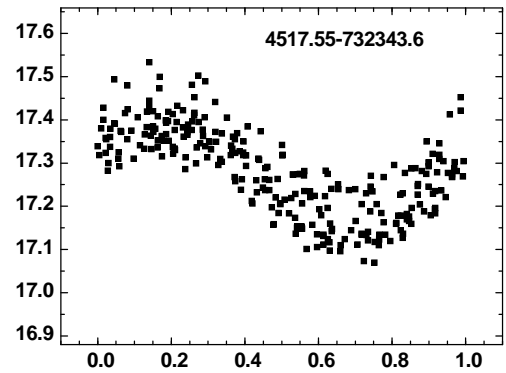
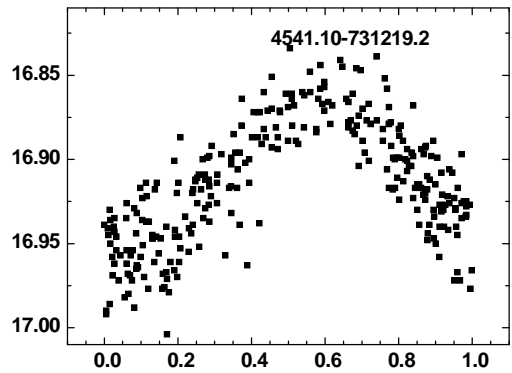
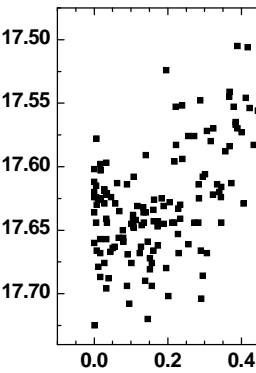
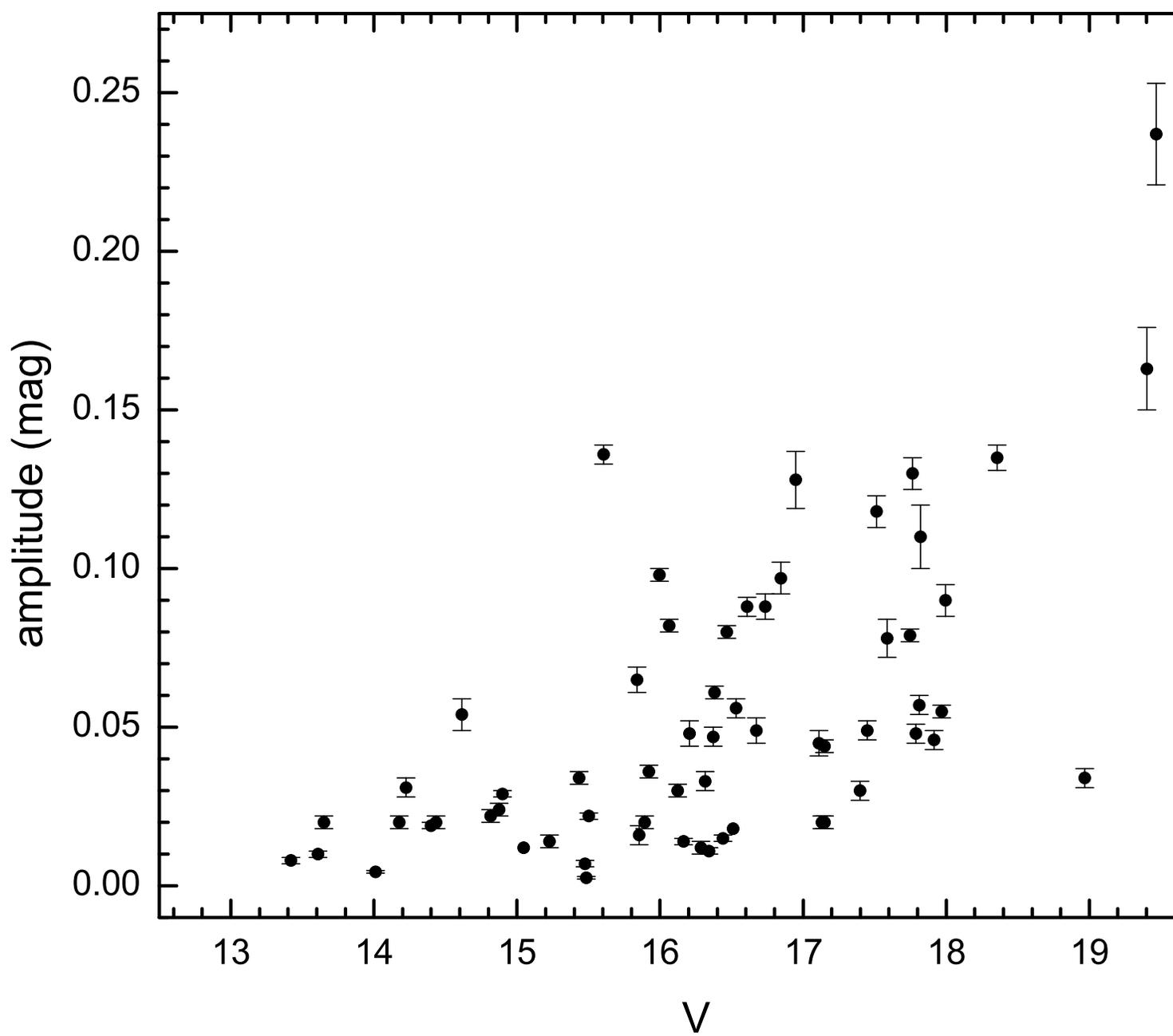
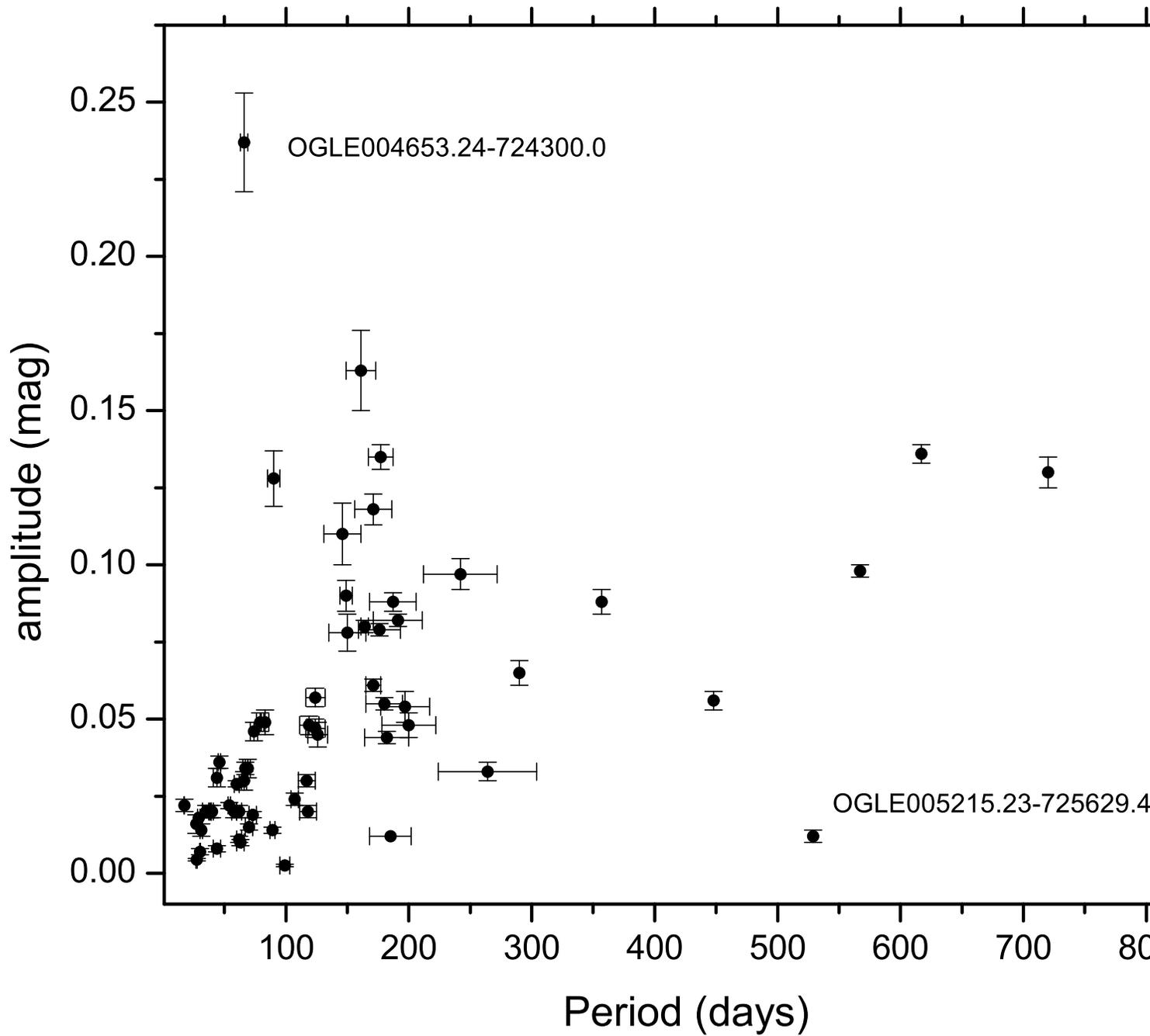
\end{document}